\documentclass[preprintnumbers,superscriptaddress,showpacs,pra]{revtex4}%
\usepackage[colorlinks=true,linkcolor=blue]{hyperref}
\usepackage{amssymb}
\usepackage{amsfonts}
\usepackage{amsmath}
\usepackage{graphicx}%
\providecommand{\U}[1]{\protect\rule{.1in}{.1in}}

\let\stdsection\section
\renewcommand\section{\nopagebreak\stdsection}
\begin{document}
\title{Fastest Frozen Temperature for a Thermodynamic System}
\author{X. Y. Zhou, Z. Q. Yang, X. R. Tang, X. Wang}
\affiliation{School for Theoretical Physics, School of Physics and Electronics, Hunan
University, Changsha, 410082, China}
\author{Q. H. Liu}
\email{quanhuiliu@gmail.com}
\affiliation{School for Theoretical Physics, School of Physics and Electronics, Hunan
University, Changsha, 410082, China}
\affiliation{Synergetic Innovation Center for Quantum Effects and Applications (SICQEA),
Hunan Normal University,Changsha 410081, China}

\begin{abstract}
For a thermodynamic system obeying both the equipartition theorem in high
temperature and the third law in low temperature, the curve showing
relationship between the specific heat and the temperature has two common
behaviors:\ it terminates at zero when the temperature is zero Kelvin and
converges to a constant value of specific heat as temperature is higher and
higher. Since it is always possible to find the characteristic temperature
$T_{C}$ to mark the excited temperature as the specific heat almost reaches
the equipartition value, it is also reasonable to find a temperature in low
temperature interval, complementary to $T_{C}$. The present study reports a
possibly universal existence of the such a temperature $\vartheta$, defined by
that at which the specific heat falls\textit{\ fastest} along with decrease of
the temperature. For the Debye model of solids, below the temperature
$\vartheta$ the Debye's law manifest itself.

\end{abstract}
\maketitle

\section{Introduction}

A thermodynamic system usually obeys two laws in opposite limits of
temperature; in high temperature the equipartition theorem works and in low
temperature the third law of thermodynamics holds. From the shape of a curve
showing relationship between the specific heat and the temperature, the
behaviors in these two limits are \textit{qualitatively} clear: in high
temperature it approaches to a constant whereas it goes over to zero during
the temperature is lowering to the zero Kelvin. In statistical mechanics we
have usually one characteristic temperature $T_{C}$, at which the specific
heat approximately takes the value given by the equipartition theorem,
implying that degrees of freedom are freely excited. However, there is no
definition of temperature complementary to $T_{C}$ to indicate a fact only a
few of degrees of freedom are thermally excited so that the third law applies.

Take the specific heat in Debye model of solids for instance, there is the
Debye frequency to denote the largest possible frequency $\omega_{D}$. The
characteristic temperature $T_{C}$ is defined by
\cite{Simon,text0,text1,text2,text3,text4}%
\begin{equation}
k_{B}T_{C}\equiv\hbar\omega_{D} \label{1}%
\end{equation}
where $k_{B}$ is the Boltzmann's constant and $\hbar$ is the Planck's
constant. When temperature $T$ reaches $T_{C}$, nearly all vibrational modes
are excited so the specific heat $c(T)$ at the $T=T_{C}$ is approximately
$1k_{B}$ (c.f. Fig. 1)%
\begin{equation}
\frac{c(T_{C})}{c(T\rightarrow\infty)}\approx0.952. \label{2}%
\end{equation}
However, once the temperature is so low that $T\approx T_{C}/6$, the Debye's
$T^{3}$ law becomes apparent (c.f. Fig. 1)
\begin{equation}
c(T)\sim T^{3},T\lesssim T_{C}/6. \label{3}%
\end{equation}
In other words, when\ $T\lesssim T_{C}/6$, the equipartition theorem breaks
down and the third law exhibits. It is quite obvious for the Debye model of
solids, at $T_{C}/6$ where the specific heat falls or rises most rapidly, c.f.
Fig. 1.

The key findings of present study are twofold: First,\ for a thermodynamic
system where the third law of thermodynamics takes effect, there is always a
temperature at which the specific heat falls or rises most rapidly. Since it
near zero Kelvin, we call it fastest frozen temperature (FFT) for the
thermodynamic system. Secondly, the FFT can be taken as a mark of temperature
at which the third law manifests itself. It must be mentioned that as a
thermodynamic system is cooling down from the high temperature to zero Kelvin,
there is a common phenomenon known as freeze-out which has been studied in,
for instance, condensed matter physics, \cite{Simon} astroparticle physics,
\cite{fout1,fout2,fout3} heavy ion collisions \cite{fout4} and statistical
mechanics. \cite{text0,text1,text2,text3,text4} The FFT under study is a
manifestation of the freeze-out from the point of view of the specific heat.

To show the possibly universal existence of the FFT, we consider three
different kinds of systems. In section II, we study FFT in detail for the
Debye model of solids. In order to explore the FFT in a broad class of system,
we examine a model system of many independent particles of the same type in
section III. The FFT is also present in a mixture of two kinds of particles,
which is given in section IV. The final section V concludes this study.

In whole paper, we will use specific heat defined by%
\begin{equation}
c=\frac{du}{dT}, \label{c}%
\end{equation}
where $u$ is mean energy over a degree of freedom, which will be precisely
defined in sections II-IV for different systems, and in our approach, all
other parameters remain unchanged except the temperature $T$. The
FFT\textit{\ }$\vartheta$ is defined by the maximum of the $dc/dt$ as
$T\rightarrow0$,
\begin{equation}
\frac{dc}{dT}\preceq\left.  \frac{dc}{dT}\right\vert _{T=\vartheta}.
\label{FFT}%
\end{equation}

\section{Fastest frozen temperature for Debye model of solids}

Debye model of solids is an elementary model to investigate the specific heat
of solids. The total internal energy $U$ is
\cite{Simon,text0,text1,text2,text3,text4}
\begin{equation}
U=U_{0}+3Nk_{B}TD(T_{C}/T)
\end{equation}
where $N$ is the number of the atoms in the solid, and $U_{0}$ is a constant,
and $D(x)$ denotes the Debye integral,%
\begin{equation}
D(x)\equiv\frac{3}{x^{3}}\int_{0}^{x}\frac{y^{3}}{e^{y}-1}dy.
\end{equation}
The mean energy $u$ in a vibrational mode is given by
\begin{equation}
u\equiv\frac{U}{3N}.
\end{equation}
In two opposite extremes the specific heats (\ref{c}) are, respectively,%
\begin{equation}
\lim_{T\rightarrow\infty}c=k_{B},\text{ and}\lim_{T\rightarrow0}c=0.
\end{equation}

The numerical calculations give both the FFT $\vartheta=0.164T_{C}\approx
T_{C}/6$ at which $c(\vartheta)\approx0.257k_{B}\approx T_{0}/4$ and
$c(T_{C})\approx0.952k_{B}\approx k_{B}=\lim_{T\rightarrow\infty}c(T)$. From
these results, we see that $\vartheta$ is really in contrast to $T_{C}$ to
mark the frozen temperature below which the Debye's $T^{3}$ law holds. The
specific heat $c(T)$ and $dc/dt$ against temperature are plotted in Fig. 1. To
note that as the temperature rises above the FFT $\vartheta$, Debye's $T^{3}$
law fails more and more seriously.

Remember that the shape of $c(T)$ curve for the Debye model shares two common
features of all $c(T)$ curves for all thermodynamic systems that satisfy both
the equipartition theorem at high temperature and the third law at low
temperature. The FFT could then be always definable as that complementary to
$T_{C}$.

\section{Fastest frozen temperature: A Theorem and Its Illustrations}

In this section, we first prove a theorem: For a system of many independent
particles of the same type, in which the energy levels of a single particle
are discrete, it has a FFT during the system is freezing and the\ FFT can be
taken as the frozen temperature. Secondly, we gives two illustrations of the theorem.

\subsection{A theorem}

Construct a model system of many independent particles of the same species, in
which the energy levels of a single particle are discrete, i.e.,
\begin{equation}
\epsilon_{n}=\epsilon_{0}<\epsilon_{1}<\epsilon_{2}<\epsilon_{3}%
<...\label{energylevels}%
\end{equation}
in which\ the spacing between $\epsilon_{1},\epsilon_{2},\epsilon_{3},...$is
negligible but $\epsilon_{1}$ is appreciably different from ground state one
$\epsilon_{0}$ which can be conveniently chosen to be zero, $\epsilon_{0}=0$,
and the density of states can be $A\epsilon^{\left(  D-2\right)  /2}$,
\cite{text0,text1,text2,text3,text4} in which $D=1,2,3,4,5,6$ is the number of
the spatial dimension, where coefficient $A$ can be set to be unity. Utilize
the Boltzmann statistics, the partition function $Z_{1}$ is with
$\beta=\left(  k_{B}T\right)  ^{-1}$ and $w_{n}$ standing for the degeneracy
in the $n$-th level,
\begin{align}
Z_{1}  & =\sum_{n=0}w_{n}\exp\left(  -\beta\epsilon_{n}\right)  \nonumber\\
& =1+\sum_{n=1}w_{n}\exp\left(  -\beta\epsilon_{n}\right)  \nonumber\\
& \approx1+A%
{\displaystyle\int\limits_{\epsilon_{1}}^{\infty}}
e^{-\beta\epsilon}\epsilon^{\left(  D-2\right)  /2}d\epsilon\nonumber\\
& =1+A\beta^{-\frac{D}{2}}\Gamma(\frac{D}{2},\epsilon_{1}\beta
)\label{partfunction}%
\end{align}
where $\Gamma(s,x)=\int_{x}^{\infty}t^{s-1}e^{-t}dt$ is the incomplete gamma
function and $\Gamma(s)=$ $\Gamma(s,0)$ is the ordinary gamma function. The
summation in the partition function can be approximately replaced by an
integral, but to get a better result, the first term or the first few terms in
the summation are carried out separately and the rest is replaced by an
integral and we convert $\sum_{n=1}w_{n}\exp\left(  -\beta\epsilon_{n}\right)
$ into an integral. This problem is analytically tractable, but the relevant
expressions are lengthy. The energy per particle and the specific heat $c$
are, respectively, determined by%
\begin{equation}
u=-\frac{\partial\ln Z_{1}}{\partial\beta},c=\frac{du}{dT}.\label{uc}%
\end{equation}
We do not explicitly show their expressions. The computations of $c(T)$ and
$dc(T)/dT$ for $D=1,2,3,4,5,$and $6$ can be easily numerically carried out,
and all $c(T)$ are similar to that given by Fig. 1. Thus, we plot $c(T)$ and
$dc(T)/dT$ for only $D=3$ in Fig. 2. It is clearly that the $c(T_{C}%
)=1.555k_{B}\approx1.5k_{B}=\lim_{T\rightarrow\infty}c(T)$\textit{\ }where the
characteristic temperature\textit{ }%
\begin{equation}
T_{C}\equiv\frac{\epsilon_{1}-\epsilon_{0}}{k_{B}}=\frac{\epsilon_{1}}{k_{B}%
}.\label{Tc}%
\end{equation}
The values of the\ FFT $\vartheta$ and ratios of two specific heats
$c(\vartheta)/c(T_{C})$ are listed in Table I.

\begin{table}[th]
\caption{$\vartheta$ and $c(\vartheta)/c(T_{C})$ for $D=1,2,3,4,5,6$}%
\centering
\par%
\begin{tabular}
[c]{|l|l|l|l|l|l|l|}\hline
& $D=1$ & $D=2$ & $D=3$ & $D=4$ & $D=5$ & $D=6$\\\hline
$\vartheta/T_{C}$ & $0.332$ & $0.363$ & $0.404$ & $0.448$ & $0.474$ &
$0.477$\\\hline
$c(\vartheta)/c(T_{C})$ & $0.312$ & $0.318$ & $0.339$ & $0.380$ & $0.421$ &
$0.464$\\\hline
\end{tabular}
\end{table}

\noindent To note that all ratios $c(\vartheta)/c(T_{C})$ are smaller than
$1$, therefore it is reasonable to take $\vartheta$ for the frozen
temperature, as we have done for Debye model of solids.

Here we give four comments on the theorem above. 

1. Replacement of the summation (\ref{partfunction}) by an integral is an
approximate treatment which is a common technique used in calculations of the
partition function in statistical mechanics. If the summation can be
explicitly carried out, the result can never be qualitatively changed. Two
illustrations will be given in next subsection.

2. Once the energy spectrum of a system is totally continuously without
bounded from upper, $Z_{1}=$ $A\Gamma(s)\beta^{-\frac{D}{2}}$ and $c=const.$
in whole range of temperature, which\ is not a physical system at low
temperature for it violates the third law of thermodynamics, beyond the scope
of present study. 

3. Once the energy spectrum of a system is continuously in a limited range,
$\epsilon\in\lbrack0,\epsilon_{\max}]$ where the $\epsilon_{\max}$ is the
largest one, the system is similar to the Debye model of solids, and the
mathematical treatment is almost the same. 

4. Once the energy spectrum of a system is partially discrete in the
inner-shells and continuously in the rest, the mathematical treatment of the
partition function is similar to (\ref{partfunction}) and the results are
similar. 

\subsection{Two illustrations of the theorem}

The theorem above applies for the vibrational degrees of freedom and
rotational ones of the diatomic gas which is dilute and interaction-free. In
this subsection, we deal with these two degrees of freedom, respectively.

Illustration 1: FFT of vibrational degrees of freedom for the diatomic gas.

The energy level is\ $\epsilon_{n}=(n+1/2)\hbar\omega,(n=0,1,2,...)$, where
$\omega$ is the vibrational frequency. The partition function is, \cite{text1}%
\begin{equation}
Z_{1}=\sum\limits_{n=0}^{\infty}e^{-\beta(n+1/2)\hbar\omega}=\frac
{e^{-\beta\hbar\omega/2}}{1-e^{-\beta\hbar\omega}}%
\end{equation}
With $T_{C}=\hbar\omega/k_{B}$, we have the mean vibrational energy $u$ and
specific heat $c$, respectively,
\begin{equation}
u=\frac{\hbar\omega}{2}+\frac{\hbar\omega}{e^{T_{C}/T}-1},c=k_{B}\left(
\frac{T_{C}}{T}\right)  ^{2}\frac{e^{T_{C}/T}}{\left(  e^{T_{C}/T}-1\right)
^{2}}.
\end{equation}
The specific heat at $T_{C}$ is $c(T_{C})$ $=0.921k_{B}\simeq\lim
_{T\rightarrow\infty}c(T)=k_{B}$. The FFT $\vartheta=0.223T_{C}$ at which
$c(\vartheta)=0.231k_{B}$ from which we see that vibrational degrees of
freedom are thermally depressed. The value $T_{C}$ of different diatomic gases
is of order $10^{3}$ K, \cite{text1} and we have then $\vartheta$ $\sim\left(
200-300\right)  $ K so the vibrational degrees of freedom at room temperature
are almost frozen. We plot the specific heat $c(T)$ and $dc/dt$ against
temperature in Fig. 3.

Illustration 2: FFT of rotational degrees of freedom for a heteronuclear
diatomic gas.

The energy level is\ $\epsilon_{l}=l(l+1)\hbar^{2}/2I,(l=0,1,2,...)$, where
$I$ is the moment of inertia. The partition function is,%
\begin{equation}
Z_{1}=\sum\limits_{l=0}^{\infty}(2l+1)e^{-\frac{\beta l(l+1)\hbar^{2}}{2I}}%
\end{equation}
Note that $T_{C}=\hbar^{2}/\left(  2Ik_{B}\right)  $,
\cite{text0,text1,text2,text3,text4} which would be $\hbar^{2}/\left(
Ik_{B}\right)  $ with use of (\ref{Tc}). The mean vibrational energy $u$ and
specific heat $c$ are, respectively, determined by%
\begin{equation}
u=\sum\limits_{l=0}^{\infty}\epsilon_{l}\rho_{l},c=\sum\limits_{l=0}^{\infty
}\epsilon_{l}\frac{d\rho_{l}}{dT}, \label{ur0}%
\end{equation}
where
\begin{equation}
\rho_{l}=\frac{(2l+1)e^{-l(l+1)T_{C}/T}}{\sum\limits_{l=0}^{\infty
}(2l+1)e^{-l(l+1)T_{C}/T}}. \label{probr0}%
\end{equation}
The numerical calculations give both $c(T_{C})$ $=1.07k_{B}\approx
\lim_{T\rightarrow\infty}c(T)=k_{B}$ and the FFT $\vartheta=0.390T_{C}$ at
which $c(\vartheta)=0.452k_{B}$. Once $T\ $decreases from $\vartheta$ the
rotational degrees of freedom is rapidly frozen out. The value of $T_{C}$, for
example, for HCl is about $15$ K, \cite{text1} we have $\vartheta=12$ K so the
rotational degrees of freedom at room temperature are freely excited. We plot
the specific heat $c(T)$ and $dc/dt$ against temperature in Fig. 4.

In this section, we deal with a system of many independent particles of the
same type, and show existence of the FFT and it can be taken as the mark of
the frozen temperature.

\section{Fastest frozen temperature for a mixture of two kinds of atom}

The theorem in section III is unable to directly apply to a mixture of
different kinds of atoms, but in fact the FFT also exists in similar manner.
To see it, we study a $1:3$ mixture of para-hydrogen and ortho-hydrogen gas
with consideration of vibrational degrees of freedom only.

The partition functions for ortho- and para-hydrogen are $Z_{1}^{ortho}$ and
$Z_{1}^{para}$, respectively \cite{text0,text1,text2,text3,text4}%
\begin{equation}
Z_{1}^{ortho}=\sum\limits_{l=1,3,5,...}(2l+1)e^{-\frac{\beta l(l+1)\hbar^{2}%
}{2I}},Z_{1}^{para}=\sum\limits_{l=0,2,4,...}(2l+1)e^{-\frac{\beta
l(l+1)\hbar^{2}}{2I}}.
\end{equation}
The mean vibrational energy $u$ and specific heat $c$ are, respectively,
determined by
\[
u=-\frac{\partial}{\partial\beta}\left(  \frac{3}{4}\ln Z_{1}^{ortho}+\frac
{1}{4}\ln Z_{1}^{para}\right)  ,c=\frac{du}{dT}.
\]
We still follow the convention of taking the rotational characteristic
temperature\textit{ }$T_{C}=\hbar^{2}/\left(  2Ik_{B}\right)  $.
\cite{text0,text1,text2,text3,text4} The numerical calculations give both the
FFT $\vartheta=1.33T_{C}$ at which $c(\vartheta)=0.305k_{B}$ and
$c(T_{C})=0.117k_{B}<<$ $k_{B}=\lim_{T\rightarrow\infty}c(T)$. From these
results, we see that $T_{C}$ is not excited temperature at all. According to
the theorem in section III, the characteristic temperature\textit{ }$T_{C}$
can only be defined for a given type of particle rather than the a mixture. It
is reasonable to define the characteristic temperature\textit{ }(\ref{Tc})
$3T_{C}=\epsilon_{2}-\epsilon_{0}$ and $5T_{C}=\epsilon_{3}-\epsilon_{1}$ for
ortho- and para-hydrogen, respectively. One can only take largest one $5T_{C}$
to specify the temperature above which the equipartition theorem can apply.
However, in our approach the FFT is independent of the choice of $T_{C}$. The
specific heat $c(T)$ and $dc/dt$ against temperature are plotted in Fig. 5.

Note that at $T=0$ K, hydrogen contains mainly para-hydrogen which is more
stable, and in general the concentration ratio of the ortho- to para-hydrogen
in thermal equilibrium is given by \cite{text1}%
\begin{equation}
r=3\frac{Z_{1}^{ortho}}{Z_{1}^{para}},
\end{equation}
from which we obtain%
\begin{equation}
r(T_{C})=7.44,r(\vartheta)=4.74,r(300)=3.01.
\end{equation}
Since at the room temperature $T\approx300$ K$=3.53T_{C}$ the ratio
$r\approx3$, and thus "the name ortho- is given to that component which
carries the larger statistical weight". \cite{text1}

\section{Conclusions and discussions}

Temperature is a fundamental quantity in physics. It has two unattainable
ends; one is the absolute zero and another is the infinitely large. If taking
$\beta=\left(  k_{B}T\right)  ^{-1}$ as an alternative definition of the
temperature $T$ instead, the new temperature $\beta$ has also two unattainable
ends. In physics, a thermodynamic system usually obeys two laws in opposite
limits of temperature, however, we have only one characteristic temperature,
at which the value of specific heat is approximately that given by the
equipartition theorem. We have no such a temperature to mark where the third
law comes to play. A well-defined temperature at which the specific heat
falls\textit{\ fastest} along with decrease of the temperature is identified,
the so-called FFT, which can be taken as the frozen temperature itself. For
Debye model of solids, FFT marks the temperature below which the the Debye's
$T^{3}$ law comes into play. For the dilute gas composed of noninteracting
atoms of the same type or not, FFT marks the temperature the system almost
gets stuck in the ground state eigenstate for the rational or vibrational
degrees of freedom.

We like to mention the following facts which can be easily demonstrated. There
are thermodynamic systems which violate the equipartition theorem in high
temperature, photon gas for instance, or violate the third law in low
temperature, classical ideal gas for instance, FFT should be used with care.
For the former, there is no $T_{C}$ and thus no FFT as well; and for the
latter, neither $T_{C}$ nor FFT is applicable. What is more, for
noninteracting Bosons, both $T_{C}$ and FFT exist but they are identical.

Therefore the present study shows not only existence of the FFT but also its
complementarity to $T_{C}$.

\begin{acknowledgments}
This work is financially supported by National Natural Science Foundation of
China under Grant No. 11675051.
\end{acknowledgments}

\begin{figure}[h]
\includegraphics[height=9cm]{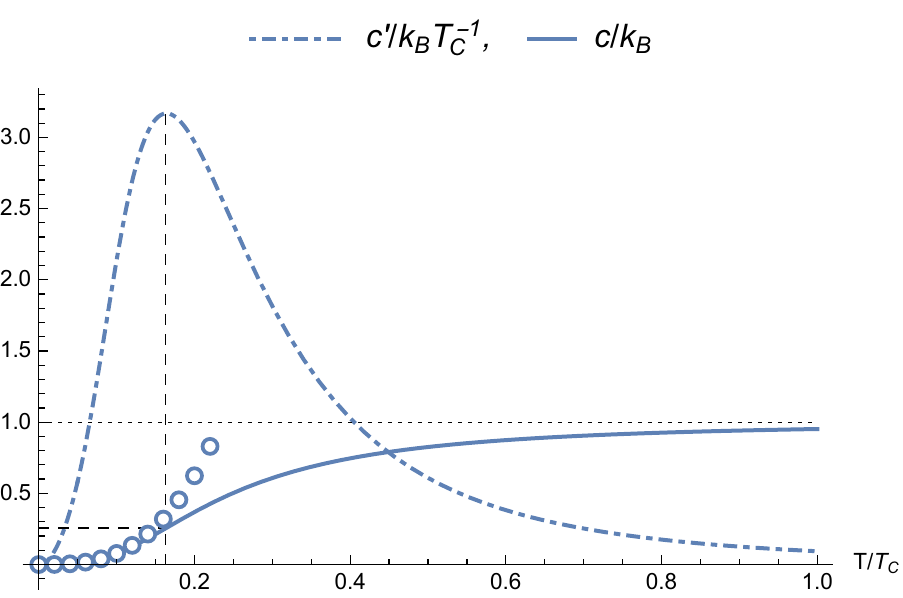}\caption{Curves for specific heat
(solid) and its derivative with respect to temperature (dashed) for the Debye
model of solid, and the auxiliary dotted lines are guide for the eye. The
equipartition value of the specific heat is $1k_{B}$, and $c(T_{C}%
)=0.952k_{B}$ $\approx$ $1k_{B}$. The $dc/dT$ has a global maximum at
$\vartheta=0.164T_{C}$ at which the specific heat $c(\vartheta)=0.257k_{B}%
\approx1/4k_{B}$. The Debye's $T^{3}$ law is plotted in circles, showing that
it holds true in temperature interval $T_{C}\in(0,\vartheta)$.}%
\label{Fig.1.}%
\end{figure}

\begin{figure}[h]
\includegraphics[height=9cm]{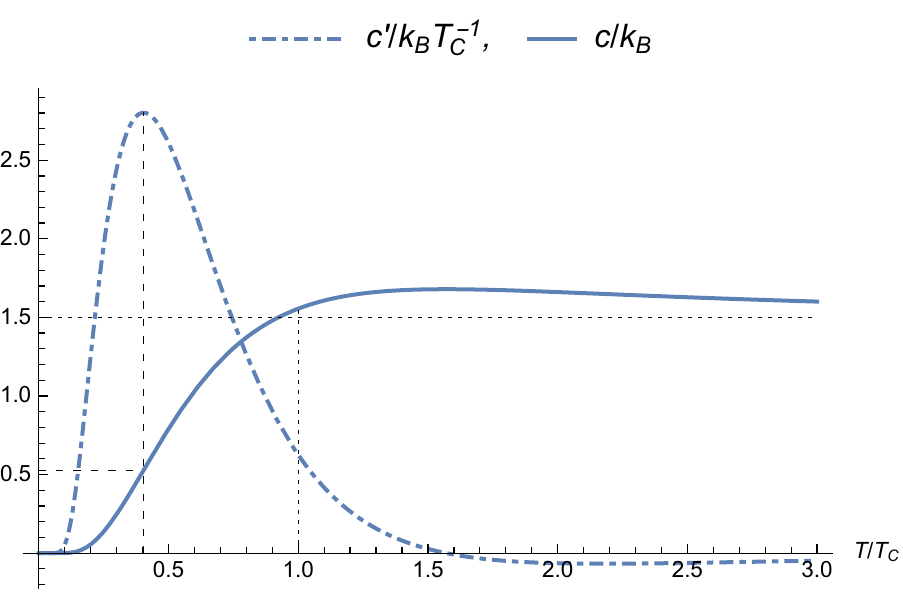}\caption{Curves for specific heat
(solid) and its derivative with respect to temperature (dashed) for our model
system, and the auxiliary dotted lines are guide for the eye. Note that the
equipartition value of the specific heat is $3k_{B}/2$, compatible with the
number of the dimension $D=3$. The $dc/dT$ has a global maximum at
$\vartheta=0.404T_{C}$ at which the specific heat $c(\vartheta)=0.527k_{B}%
=0.339c(T_{C})$. $c(T_{C})=1.56k_{B}\approx1.5k_{B}$, the equipartition value.
}%
\label{Fig.2.}%
\end{figure}

\begin{figure}[h]
\includegraphics[height=9cm]{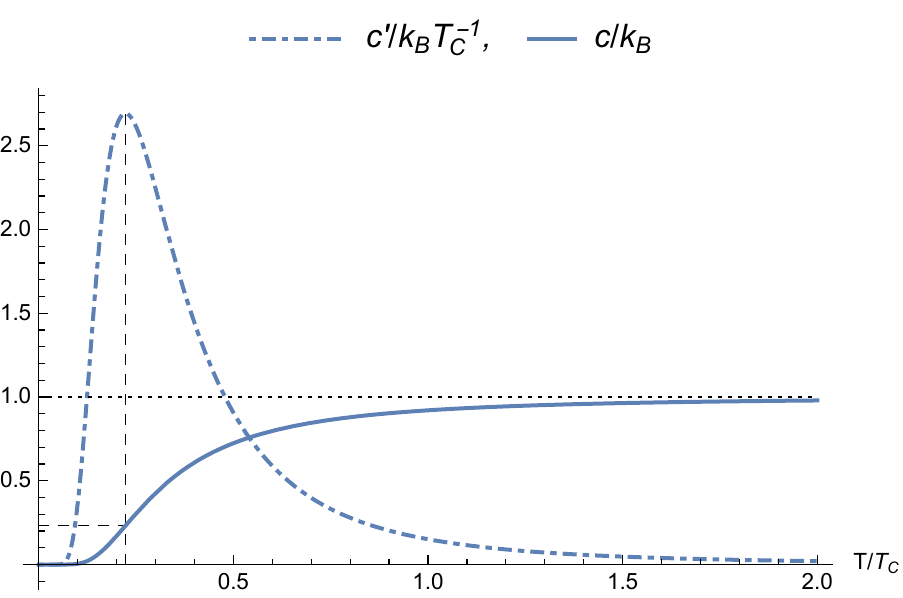}\caption{Curves for specific heat
(solid) and its derivative with respect to temperature (dashed) for the
vibrational degrees of freedom of a diatomic gas, and the auxiliary dotted
lines are guide for the eye. Note that the equipartition value of the specific
heat is $1k_{B}$, and $c(T_{C})=0.921k_{B}$ is slightly smaller than $1k_{B}$.
The $dc/dT$ has a global maximum at $\vartheta=0.223T_{C}$ at which the
specific heat $c(\vartheta)=0.231k_{B}\approx1/4k_{B}$.}%
\label{Fig.3.}%
\end{figure}

\begin{figure}[h]
\includegraphics[height=9cm]{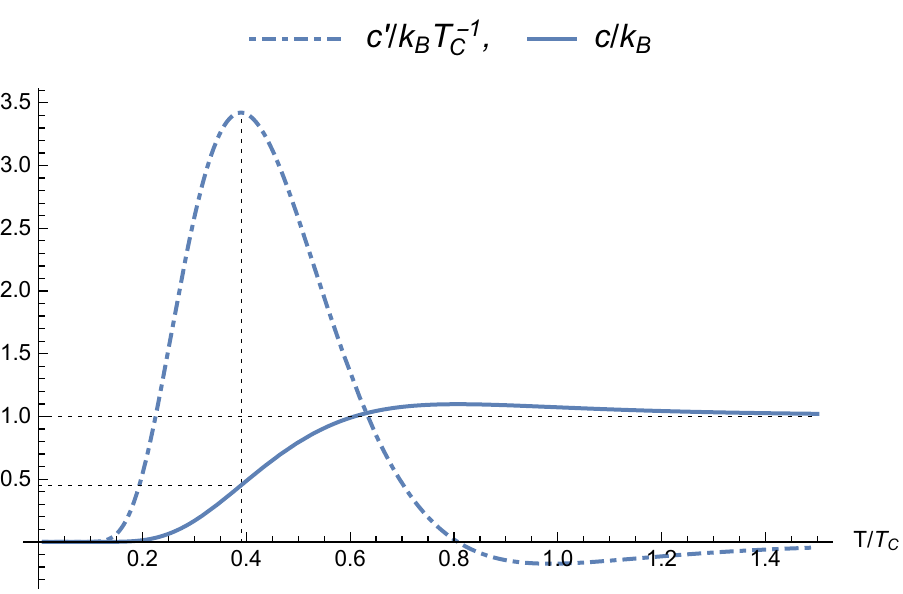}\caption{Curves for specific heat
(solid) and its derivative with respect to temperature (dashed) for the
rotational degrees of freedom of a heteronuclear diatomic gas, and the
auxiliary dotted lines are guide for the eye. Note that the equipartition
value of the specific heat is $1k_{B}$, and $c(T_{C})=1.07k_{B}\approx1k_{B}$.
The $dc/dT$ has a global maximum at $\vartheta=0.390T_{C}$ at which the
specific heat $c(\vartheta)=0.452k_{B}=0.421c(T_{C})$.}%
\label{Fig.4.}%
\end{figure}

\begin{figure}[h]
\includegraphics[height=9cm]{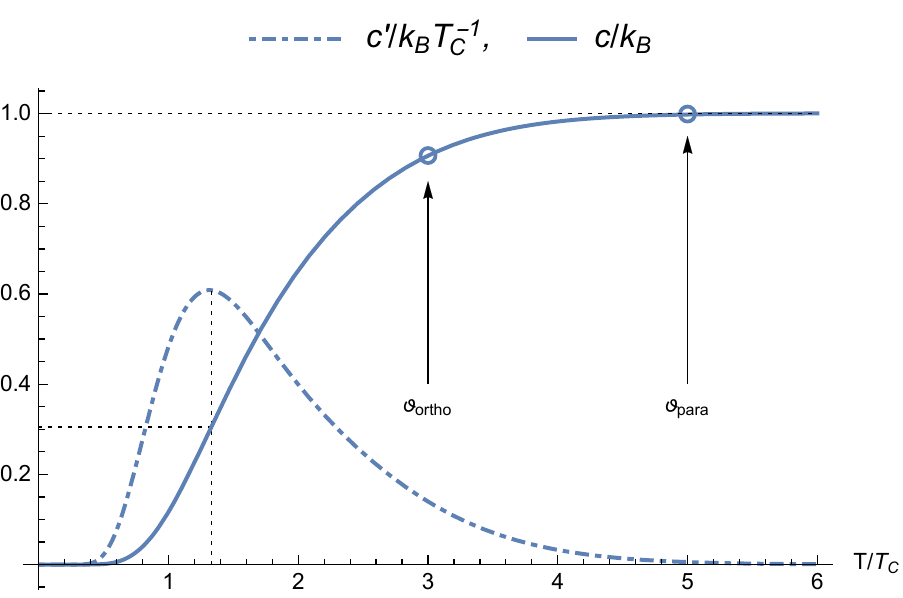}\caption{Curves for specific heat
(solid) and its derivative with respect to temperature (dashed) for the
rotational degrees of freedom of $1:3$ mixture of para-hydrogen and
ortho-hydrogen molecules, and the auxiliary dotted lines and arrows are guide
for the eye. The equipartition value of the specific heat is $1k_{B}$, and
$c(T_{C})=0.117k_{B}$ is much smaller than $1k_{B}$. The $dc/dT$ has a global
maximum at $\vartheta=1.33T_{C}$ at which the specific heat $c(\vartheta
)=0.305k_{B}$. The conventional definition of characteristic temperature
$T_{C}=\hbar^{2}/(2Ik_{B})$ is nothing but a unit, and once we use our
definition (\ref{Tc}), we have for ortho- and para-hydrogen, respectively,
$c(\vartheta_{\text{ortho}})=0.907k_{B}\approx1k_{B}$ and $c(\vartheta
_{\text{para}})=0.999k_{B}\approx1k_{B}$.}%
\label{Fig.5.}%
\end{figure}
\end{document}